\documentclass[prl,twocolumn,floatfix,amsmath,nofootinbib,amssymb,floatfix]{revtex4}
\usepackage{graphicx,color,dcolumn,booktabs,bm}
\usepackage{longtable,lscape}
\usepackage{txfonts}
\usepackage{overpic}
\usepackage{amssymb}
\usepackage{indentfirst}
\usepackage{feynmf}   
\usepackage{slashed}  
\usepackage{cases}
\usepackage{color}
\usepackage{multirow}
\usepackage{epstopdf}
\usepackage{enumerate}
\usepackage{graphicx,color,dcolumn,booktabs,bm}
\usepackage[colorlinks, citecolor=blue,anchorcolor=red,menucolor=red, linkcolor=red,filecolor=red,urlcolor=blue,frenchlinks=red]{hyperref}

\usepackage{orcidlink}

\begin{document}

\title{Cut structures and an observable singularity in the three-body threshold dynamics: the $T_{cc}^+$ case}
\author{Jun-Zhang Wang\,\orcidlink{0000-0002-3404-8569}}\email{wangjzh2022@pku.edu.cn}
\author{Zi-Yang Lin\,\orcidlink{0000-0001-7887-9391}}\email{lzy$_$15@pku.edu.cn}
\author{Shi-Lin Zhu\,\orcidlink{0000-0002-4055-6906}}\email{zhusl@pku.edu.cn
}
\affiliation{School of Physics and Center of High Energy Physics, Peking University, Beijing 100871, China}

\date{\today}

\begin{abstract}
The three-body threshold effect, the distinctive and intriguing non-perturbative dynamics in the low-energy hadron-hadron scattering, has acquired compelling significance in the wake of the recent observation of the double-charm tetraquark $T_{cc}^+$.  This dynamics is characterized by the emergence of singular points and branch cuts within the interaction potential, occurring when the on-shell condition of the mediated particle is satisfied. The presence of these potential singularities indicates that the system is no longer Hermitian and also poses intractable challenges in obtaining exact solutions for dynamical scattering amplitudes. In this work, we develop a complex scaled Lippmann-Schwinger equation as an operation of analytical continuation of the $T$ matrix to resolve this problem. Through a practical application to the $DD^* \to DD^*$ process, we reveal complicated cut structures of the three-body threshold dynamics in the complex plane, primarily stemming from the one-pion exchange. Notably, our methodology succeeds in reproducing the $T_{cc}^+$ structure, in alignment with the quasi-bound pole derived from the complex scaling method within the Schrödinger equation framework. More remarkably, after solving the on-shell $T$ matrix on the positive real axis of momentum plane, we find an extra new structure in the $DD^*$ mass spectrum, which arises from a right-hand cut at a physical pion mass and should be observable in Lattice QCD simulations and future high-energy experiments. 
\end{abstract}

\maketitle

\section{Introduction}
Low-energy hadron-hadron interactions offer a window into the non-perturbative dynamics of the fundamental theory of the strong force, i.e., Quantum Chromodynamics (QCD), which has been one of the most important issues in particle physics and nuclear physics. Due to the color confinement of QCD, theorists have come to realize that hadrons can be treated as an effective basic freedom in the low-energy strong interaction. Thus, a modern advanced tool of effective field theory has been proposed to describe these interactions, such as the Chiral Perturbation Theory (ChPT) based on the spontaneous breaking of chiral symmetry in QCD \cite{Weinberg:1978kz,Gasser:1983yg,Gasser:1984gg,Weinberg:1990rz,Weinberg:1991um,Jenkins:1990jv,Bernard:1992qa,Hemmert:1997ye,Epelbaum:2008ga,Machleidt:2011zz,Meng:2022ozq}. In the framework of the effective field theory, the strong force is mediated by the exchange of mesons, particularly pions. A representative example is the nucleon-nucleon interaction, the attractive and repulsive behaviors of which are pivotal for comprehending nuclear forces and atomic nucleus properties \cite{Epelbaum:2008ga,Machleidt:2011zz}. 

The exciting advancement in the realm of low-energy strong interactions still continues. Very recently, the LHCb Collaboration observed a double-charm exotic hadron  $T_{cc}^+$ in the mass spectrum of $D^0D^0\pi^+$ \cite{LHCb:2021vvq,LHCb:2021auc}, in which the extracted pole information is given by 
\begin{eqnarray}
    \delta m_{\mathrm{pole}}=-360\pm40~ \mathrm{keV}, ~~ \Gamma_{\mathrm{pole}}=48\pm2~ \mathrm{keV},
\end{eqnarray}
with a unitarized Breit-Wigner parameterization scheme \cite{LHCb:2021auc}. Here, $\delta m=m-m_{D^0}-m_{D^{*+}}$. It is evident that the pole position of $T_{cc}^+$ lies in extremely close proximity to the $D^0D^{*+}$ threshold. Consequently, this state has commonly been regarded as a good candidate of a heavy-flavored hadronic molecule, formed through the interaction of charmed mesons $DD^*$ \cite{Manohar:1992nd,Janc:2004qn,Ohkoda:2012hv,Li:2012ss,Chen:2021cfl,Dong:2021bvy,Feijoo:2021ppq,Albaladejo:2021vln,Fleming:2021wmk,Meng:2021jnw,Du:2021zzh,Lin:2022wmj,Cheng:2022qcm,Ke:2021rxd,Ling:2021bir,Liu:2019yye,Yan:2021wdl,Jin:2021cxj,Xin:2021wcr,Shi:2022slq,Ortega:2022efc,Du:2023hlu,Wang:2022jop,Chen:2023fgl,Chen:2021vhg}. Undoubtedly, the discovery of $T_{cc}^+$ offers an exceptional opportunity to illuminate the intricate internal dynamics of the hadron-hadron interactions involving heavy quarks.

Similar to the nucleon-nucleon interactions, the heavy-meson-heavy-meson interactions share a common underlying dynamics characterized by the exchange of pions and heavier isoscalar mesons. However, a unique aspect of the $DD^*$ interaction is the inherent instability of the $D^*$ meson, leading to its decay into $D\pi$ and the possibility of introducing an on-shell intermediate pion meson in the pion-exchange interactions. At the leading order, this on-shell singularity appearing in the one-pion-exchange (OPE) potential, results in a non-vanishing imaginary component and renders the Hamiltonian non-Hermitian. According to the optical theorem, this imaginary part of the OPE potential corresponds to the three-body $DD\pi$ decay. Importantly, the final $DD\pi$ states is the sole strong decay mode of $T_{cc}^+$. This underscores the critical role of the three-body dynamics in unveiling the nature of the special $T_{cc}^+$ state. In previous works \cite{Lin:2022wmj,Cheng:2022qcm}, a revised OPE potential incorporating the three-body threshold dynamics within a relativistic pion propagator has been achieved, which effectively explains the observed narrow width of the $T_{cc}^+$ state.

In the context of the OPE potential involving the three-body dynamics, in addition to a unitary cut at the three-body threshold, more plentiful cut structures exist, which potentially give rise to intriguing physical phenomena. 
In this work, we systematically study the cut structures of the three-body threshold dynamics in the complex plane, which include the cases of the on-shell amplitudes and half-on-shell amplitudes when opening or closing the three-body dynamics. Subsequently, we concentrate on the situation of interest to us in which the three-body threshold dynamics is active, and find that the branch cut of the half-on-shell amplitude with an imaginary on-shell momentum always traverses the positive real axis, which may lead to unavailability of the conventional Lippmann-Schwinger equation in obtaining the physical on-shell $T$ matrix below the energy threshold. 
In order to resolve this problem, we develop a complex scaled Lippmann-Schwinger equation (CSLSE) approach, which ensures a logical treatment of analytical continuation of the $T$ matrix. A most crucial finding derived from the CSLSE emphasizes the necessity of shifting the integral path of the loop momentum when studying the physical on-shell $T$ matrix below the energy threshold. Rather than following the default physical real axis, this path should now be close to the negative imaginary axis. Additionally, the another advantage of this method is its high efficiency in addressing the divergence issue associated with the double singular points of the potential function along the integral path when computing the physical on-shell $T$ matrix above the energy threshold. This singularity divergence is often not straightforward to solve using the conventional Lippmann-Schwinger equation.  This approach should be universally applicable for studying this type of interaction encompassing the three-body effect and more generalized one-boson exchange potential. By taking the isoscalar $DD^*$ scattering as an example, the observed quasi-bound $T_{cc}^+$ structure in LHCb can be clearly reproduced in the CSLSE.  
More intriguingly, we find an extra new structure in the $DD^*$ mass spectrum for the first time by solving the physical on-shell $T$ matrix above the energy threshold. This structure arises from the right-hand cut of the OPE potential involving the three-body dynamics, which should be regarded as a distinctive symbol for assessing the role of the three-body dynamics in governing the heavy-hadron-heavy-hadron system.

\section{Cut structures in the analytical extension of the three body $DD\pi$ dynamics}

\begin{figure}[t]
\centerline{\includegraphics[width=8.7cm]{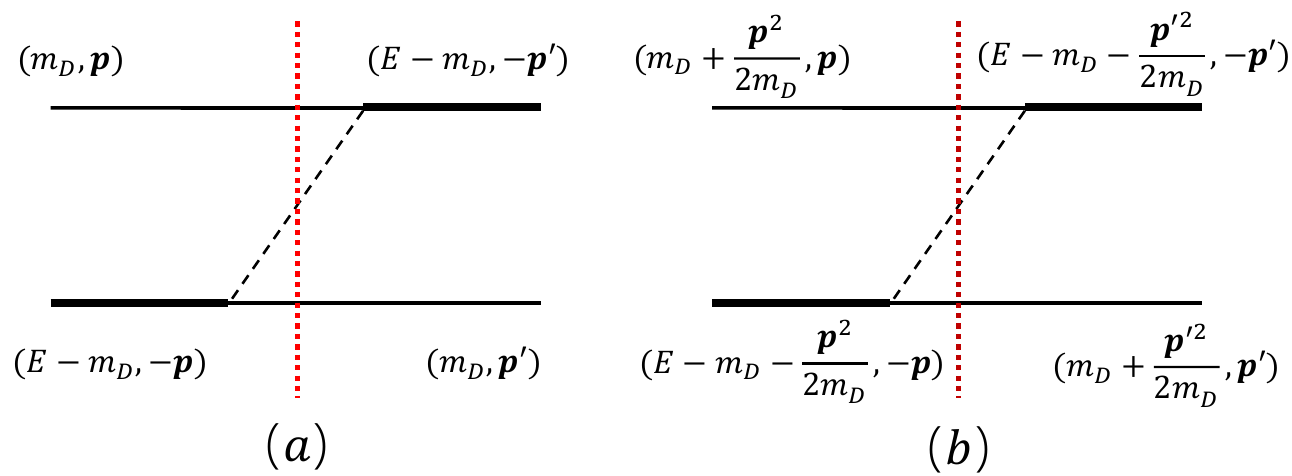}}
	\caption{Three-body $DD\pi$ intermediate states in the OPE potential without and with the kinetic energy terms of the heavy mesons as shown in figure $(a)$ and $(b)$, respectively. \label{fig1} }
\end{figure}

In the heavy meson chiral effective field theory (HMChEFT) \cite{Meng:2022ozq}, the leading order interactions of the $DD^* \to DD^*$ scattering include the contact term and OPE potential. In the isospin limit,  the interaction of the isoscalar $DD^*$ state $(I=0)$  can be written as
\begin{eqnarray}
 &&V^{I=0}(p,p^{\prime},z)=C_t-\frac{3g^2}{8f_{\pi}^2}\frac{(\varepsilon \cdot q)(\varepsilon^{\prime} \cdot q)}{q^2-m_{\pi}^2+i\epsilon} \nonumber \\
 &&=C_t-\frac{g^2}{8f_{\pi}^2}\frac{3(\varepsilon \cdot q)(\varepsilon^{\prime} \cdot q)}{q_0^2-(p^2+p^{\prime 2}-2pp^{\prime} z)-m_{\pi}^2+i\epsilon}. 
\end{eqnarray}
The usual instantaneous approximation of $q_0=0$ will lead to the OPE potential without singularity. Here, $q_0\sim(m_{D^*}-m_D)$ is comparable with the mass $m_{\pi}$, so the instantaneous approximation is not appropriate. In addition, the fact of $q_0\sim(m_{D^*}-m_D)>m_{\pi}^{\mathrm{phy}}$ leads to an on-shell pion exchange and then the $DD^*$ system may decay to the three-body channel of $DD\pi$. Since we consider only the S-wave interactions, in order to introduce this three-body threshold dynamics (see Fig. \ref{fig1}), the OPE potential can be re-written as
\begin{eqnarray}
V^{I=0}_{\mathrm{OPE}}=-\frac{g^2}{8f_{\pi}^2}\frac{(p^2+p^{\prime 2}-2pp^{\prime} z)(\varepsilon^{\prime} \cdot \varepsilon)}{(E^{\prime}+\delta)^2-(p^2+p^{\prime 2}-2pp^{\prime} z)-m_{\pi}^2+i\epsilon}, \label{eq3}
\end{eqnarray}
where $E^{\prime}=k_0^2/2\mu$ and $\delta=m_{D^*}-m_{D}$. Here, for the concerned physical quantities near the threshold, ignoring the kinetic energy of the heavy meson is obviously a good approximation. If considering the kinetic energy term, $\delta=m_{D^*}-m_{D}-(p^2+p^{\prime2})/(2m_{D})$.  We further define an effective mass square $m_{eff}^2=(E^{\prime}+\delta)^2-m_{\pi}^2$. For the $S$-wave scattering process, its partial-wave-projected components can be obtained by

\begin{eqnarray}
V^{I=0}_{S}(p,p^{\prime})=4\pi C_t+\int_{-1}^{1}dz 2\pi V^{I=0}_{\mathrm{OPE}}(p,p^{\prime},z).  \label{eq4}
\end{eqnarray}

As the scattering amplitude of the leading-order Born approximation, $V^{I=0}_{S}(p,p^{\prime})$, does not satisfy the unitarity. In order to ensure the unitarity and produce bound states, resonances or virtual states \cite{Chen:2023eri}, the re-summation via a dynamical equation is needed. For the two-particle system, its dynamical scattering can be described by the Lippmann-Schwinger equation (LSE) or Schrödinger equation. The LSE is given by
\begin{eqnarray}
  T_{\alpha\beta}(p,p^{\prime},k_0)&\bm{=}&V_{\alpha\beta}(p,p^{\prime},k_0)+\sum_{\gamma}\int_0^{\infty}\frac{dq q^2}{(2\pi)^3}V_{\alpha\gamma}(p,q,k_0)  \nonumber \\ 
  && \times G_{\gamma}(q,k_0) T_{\gamma\beta}(q,p^{\prime},k_0),
\end{eqnarray}
with
\begin{eqnarray}
 G_{\gamma}(q,k_0)\bm{=}\frac{2\mu_{\gamma}}{k_0^2-q^2+i\epsilon},
\end{eqnarray}
where $V_{\alpha\beta}(p,p^{\prime},k_0)$ is the partial-wave-projected potential of the $\alpha$ channel to the $\beta$ channel, and $p$ and $p^{\prime}$ correspond to the initial and final momentum, respectively. The center-of-mass momentum $k_0$  and the reduced mass $\mu_{\gamma}$ are defined by
\begin{eqnarray}
 k_0^2\bm{=}2\mu_{\gamma}(E-m_1-m_2), ~~~\mu_{\gamma}\bm{=}\frac{m_1m_2}{m_1+m_2}.
\end{eqnarray}
Considering a single channel case $DD^* \to DD^*$ ($I=0$), the Green's function in the dynamical equation includes a right-hand unitary cut from the two-body threshold, i.e., $k_0^{\mathrm{Rh}}=\sqrt{2\mu(m_D+m_{D^*})}$.

\begin{figure*}[t]
\centerline{\includegraphics[width=15.0cm]{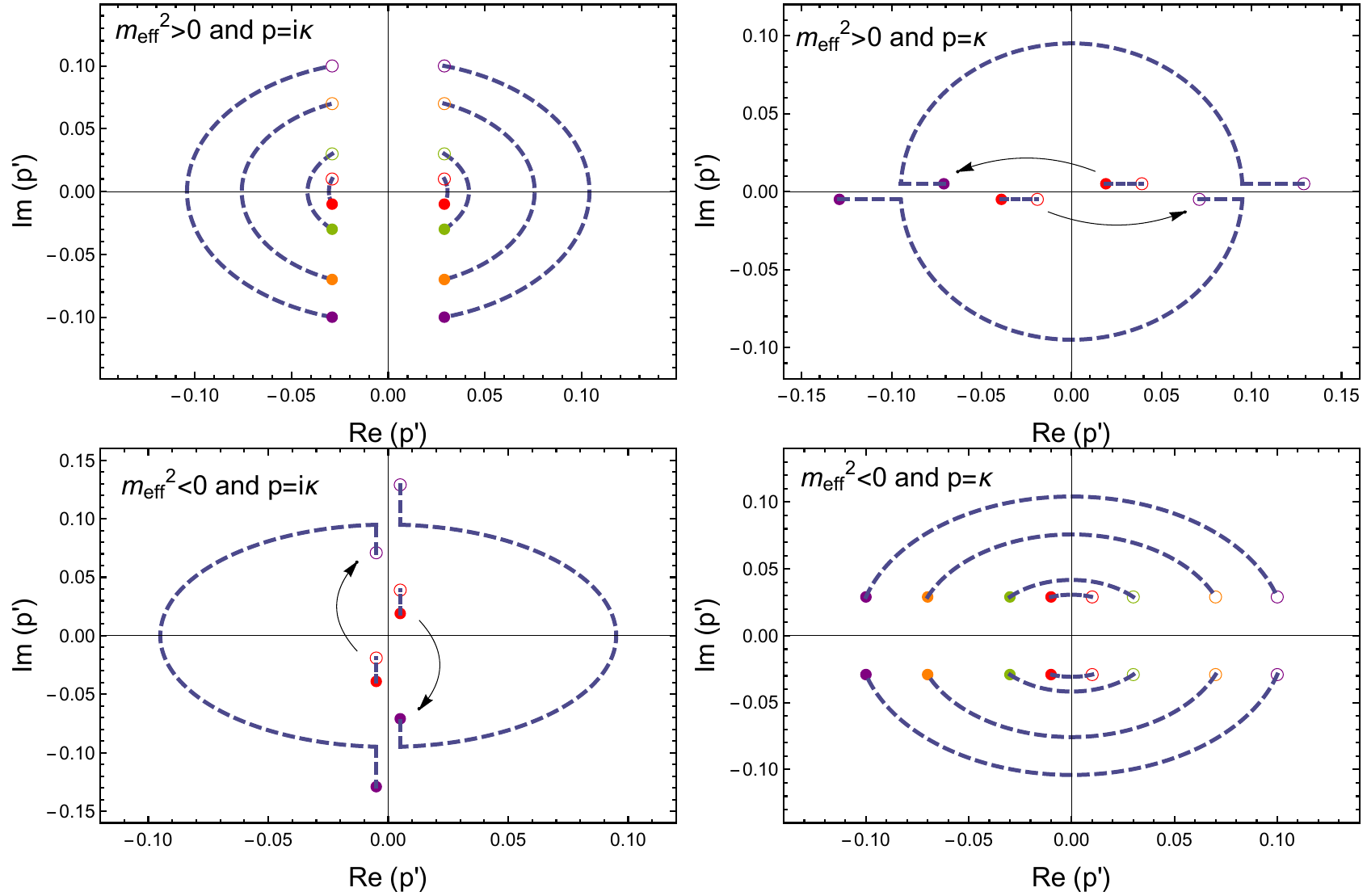}}
	\caption{ The cut structures of the half-on-shell scattering amplitudes from the OPE potential involving the three-body dynamics under different cases. The red, green, orange and purple singularities correspond to $\text{\large{$\kappa$}}=0.01$, 0.03, 0.07 and 0.1 GeV, respectively and the dashed line denotes the branch cuts. \label{fig2} }
\end{figure*}

For the OPE potential involving the three-body dynamics, its cut structures become more complicated. The revised OPE potential 
includes a basic three-body threshold cut. It can be seen from Eq. (\ref{eq3}) that when $p=p^{\prime}=0$
\begin{eqnarray}
    V^{I=0}_{\mathrm{OPE}}(p,p^{\prime})&\propto&((k_0^2/(2\mu)+\delta+m_{\pi})((k_0^2/(2\mu)+\delta-m_{\pi}))^{-1} \\
    &\propto&((E-m_D-m_D-m_{\pi})(E-m_D-m_D+m_{\pi}))^{-1}, \nonumber
\end{eqnarray}
 where the three-body cut appears at $E=m_D+m_D+m_{\pi}$ or $k_0^{\mathrm{Rh}}=\sqrt{2\mu(m_{\pi}-\delta)}$. In addition, for the on-shell scattering amplitude ($p=p^{\prime}=k_0$), which can be directly detected in experiments, the partial-wave OPE potential involving the three-body dynamics will induce a new cut structure on the real axis of the energy plane. The position of this branch point can be derived from Eq. (\ref{eq4}), i.e., 
\begin{eqnarray*}
(k_0)^2=m_{eff}^2/4=((E^{\prime}+\delta)^2-m_{\pi}^2))/4.
\end{eqnarray*}
When $m_{eff}^2<0$, which could correspond to a larger unphysical pion mass, the three-body threshold is not enabled. This cut becomes a left-hand cut below the energy threshold. However, for a physical $m_{\pi}$ ($m_{eff}^2>0$), it can be expected that there exists a right-hand cut above the $DD^*$ threshold, which is different from the unitary cut at the $DD\pi$ threshold.

In fact, we can further make an analytical continuation of the potential $V^{I=0}_{S}(p,p^{\prime})$ to study the cut structures of the three-body dynamics in the complex plane, which is usually related to the half-on-shell scattering amplitude ($k_0=p$). The behaviors of their singularities (branch points) and branch cuts are obviously distinct according to the signs of $k_0^2$ and $m_{eff}^2$, which are shown in Fig. \ref{fig2} by taking the constant $m_{eff}^2=\pm8.43\times10^{-4}$ GeV$^2$ (ignore the dependence of $k_0$ in $m_{eff}^2$ ) as an example. Here, this absolute value of $m_{eff}^2$ is obtained by the input of $m_{D}=1.867, m_{D^*}=2.009$, $m_{\pi}=0.139$ GeV. The main contents and conclusions of four cases are summarized

\begin{itemize}
    \item For the case with the physical pion mass ($m_{eff}^2>0$) and center-of-mass energy $E_{\textrm{cm}}$ below the threshold ($k_0=p=i\text{\large{$\kappa$}}$ with real $\text{\large{$\kappa$}}$), it has four branch points $(m_{eff}+i\text{\large{$\kappa$}})$, $(m_{eff}-i\text{\large{$\kappa$}})$, $(-m_{eff}+i\text{\large{$\kappa$}})$ and $(-m_{eff}-i\text{\large{$\kappa$}})$ in the complex plane of $p^{\prime}$. It can be found that the branch cuts are transversely symmetrical and the path along the real axis will encounter branch cuts.
    \item For the case with the physical pion mass and $E_{\textrm{cm}}$ above the threshold ($k_0=p=\text{\large{$\kappa$}}$ with real $\text{\large{$\kappa$}}$), it has four branch points $(\text{\large{$\kappa$}}+m_{eff}+i\epsilon)$, $(\text{\large{$\kappa$}}-m_{eff}-i\epsilon)$, $(-\text{\large{$\kappa$}}+m_{eff}+i\epsilon)$ and $(-\text{\large{$\kappa$}}-m_{eff}-i\epsilon)$ on the real axis. When $\text{\large{$\kappa$}}$ is relatively small, its branch cut is a line segment.  However, an interesting phenomenon is that each of the two branch points near the origin crosses the imaginary axis and moves to the opposite half-plane as $\text{\large{$\kappa$}}$ gradually increasing till a critical point $\text{\large{$\kappa$}}=|m_{eff}|$. At this point the branch cut will change drastically, whose closed behavior covers almost every possible path from the origin. It is worth noting that the real axis is the only path which does not encounter branch cuts after considering an infinitesimal imaginary part $i\epsilon$.
    \item For the case with the unphysical pion mass ($m_{eff}^2<0$) and $E_{\textrm{cm}}$ below the threshold, it has four branch points $((\text{\large{$\kappa$}}+m_{eff})i+\epsilon)$, $((\text{\large{$\kappa$}}-m_{eff})i-\epsilon)$, $((-\text{\large{$\kappa$}}+m_{eff})i+\epsilon)$ and $((-\text{\large{$\kappa$}}-m_{eff})i-\epsilon)$ on the imaginary axis. As $\text{\large{$\kappa$}}$ increases, the behaviors of the branch points are similar to the above case. The only difference is that they are aligned along the imaginary axis. The imaginary axis is the only path which  will not encounter branch cuts when $\text{\large{$\kappa$}}>|m_{eff}|$. 
    \item For the case with the unphysical pion mass and $E_{\textrm{cm}}$ above the threshold, which is also resemblance to the generalized one-boson exchange potential with a heavier mediated meson, it has four branch points $(\text{\large{$\kappa$}}+im_{eff})$, $(\text{\large{$\kappa$}}-im_{eff})$, $(-\text{\large{$\kappa$}}+im_{eff})$ and $(-\text{\large{$\kappa$}}-im_{eff})$. It can be found that the branch cuts are longitudinally symmetrical and the path along the real axis will not encounter branch cuts.
\end{itemize}

\begin{figure}[t]
\centerline{\includegraphics[width=7.5cm]{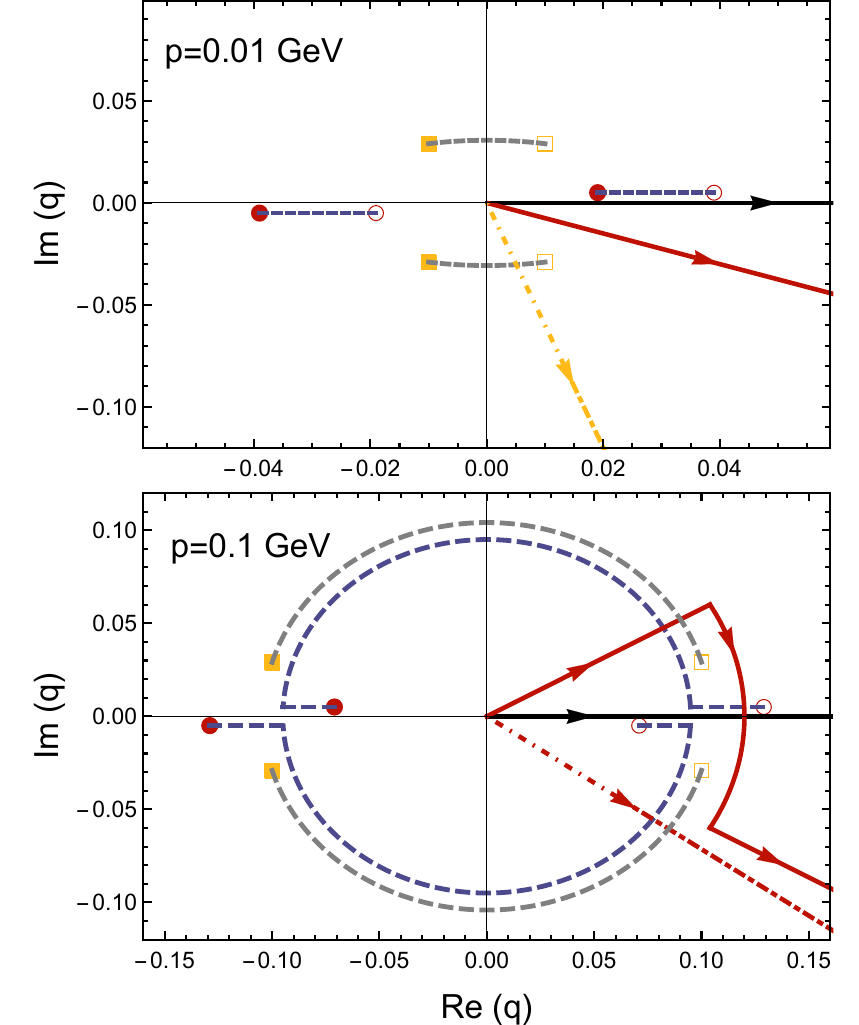}}
	\caption{Integral path schemes in the complex plane for the solution of the on-shell $T$ matrix with a positive real momentum. The dot dashed line denotes that the path will encounter the branch cut. The red and yellow points and lines  correspond to the case with $m_{eff}^2>0$ and $m_{eff}^2<0$, respectively.  \label{fig3} }
\end{figure}

\section{The complex scaled Lippmann-Schwinger equation}

Due to the intricate cut structures arising from the three-body $DD\pi$ effect, its dynamical re-summation within the framework of LSE is also expected to be more complicated.
The conventional Lippmann-Schwinger equation is
\begin{eqnarray}
T(p,p^{\prime},k_0)&\bm{=}&V(p,p^{\prime},k_0)+\int_0^{\infty}\frac{dq q^2}{(2\pi)^3}V(p,q,k_0)   \nonumber \\
&&\times G(q,k_0) T(q,p^{\prime},k_0), \label{eq9}
\end{eqnarray}
where the integral path along the loop momentum $q$ follows a physical positive real axis. For the solution of the on-shell $T$ matrix with $E_{cm}$ below the threshold and physical pion mass, as depicted in Fig. \ref{fig2}, it is evident that the path along real axis always intersects the branch cut of the potential. Such a phenomenon does not happen in the case of the on-shell $T$ matrix with $E_{cm}$ above the threshold. Additionally, for $V(p,q,k_0)=V^{I=0}_{S,\mathrm{OPE}}(\text{\large{$\kappa$}},q,\text{\large{$\kappa$}})$ with the physical pion mass, as shown in Fig. \ref{fig2}, it introduces two singular points precisely on the integral path along the positive real axis, and the resulting divergence poses challenges in accurately computing the on-shell amplitude $T(\text{\large{$\kappa$}},\text{\large{$\kappa$}},\text{\large{$\kappa$}})$ using the dynamical equation. Although it is still feasible to directly encompass the contribution of this divergence using an alternative scheme within the conventional LSE framework (as detailed in Appendix A), such a formalism demands sufficiently high numerical accuracy.

In this work, we develop a complex scaled Lippmann-Schwinger equation to effectively address these issues and challenges brought about by the cut structures of the three-body dynamics. One significant advantage of this approach lies in its remarkable efficiency in resolving the divergence issue associated with the singular points in both the potential and Green's function.  In order to avoid the singularities of the potential, one can extend the integral path of the LSE to the complex plane to ensure an identical integral result. In Fig. \ref{fig3}, we propose a set of path schemes in complex space for solving the physical on-shell $T$ amplitude above the energy threshold, which are different in terms of the magnitude of momentum $p$. Let us begin with the case involving a physical pion mass. When $p=\text{\large{$\kappa$}}$ is not larger than the critical point $\text{\large{$\kappa$}}=|m_{eff}|$, the branch cut locates on real axis. Then a complex scaling rotation transformation $q \to qe^{-i\theta}$ is introduced and the LSE becomes 
\begin{eqnarray}
&&T(p,p^{\prime},k_0)\bm{=}V(p,p^{\prime},k_0)+\int_0^{\infty}\frac{d^3\bm{q}}{(2\pi)^3}e^{-i3\theta}\nonumber \\
&&~~~~~~~~~~~~V(p,qe^{-i\theta},k_0)G_{\gamma}(qe^{-i\theta},k_0) T(qe^{-i\theta},p^{\prime},k_0). \label{eq10}
\end{eqnarray}
However, for $p>|m_{eff}|$, this path will cross the circular branch cut and enter another Riemann surface of the potential function and thus is no longer applicable. Here, an available way is to first go through the circular branch cut in the upper half-plane and then re-cross the same branch cut back to the lower half-plane, whose feasibility can be assured by Cauchy integral theorem. On the unphysical pion mass conditions, the situation becomes more manageable because the singular points of the potential are distant from the real axis (as shown in Fig. \ref{fig3}). Moreover, the complex scaled path remains applicable and offers a distinct benefit by automatically incorporating the discontinuity component from the Green's function. It is worth emphasizing that the LSE in Eq. (\ref{eq10}) is no longer an iterative equation and cannot be directly solved. Therefore, a key issue revolves around the analytical continuation of the $T$ matrix.

\begin{figure}[t]
\centerline{\includegraphics[width=8.8cm]{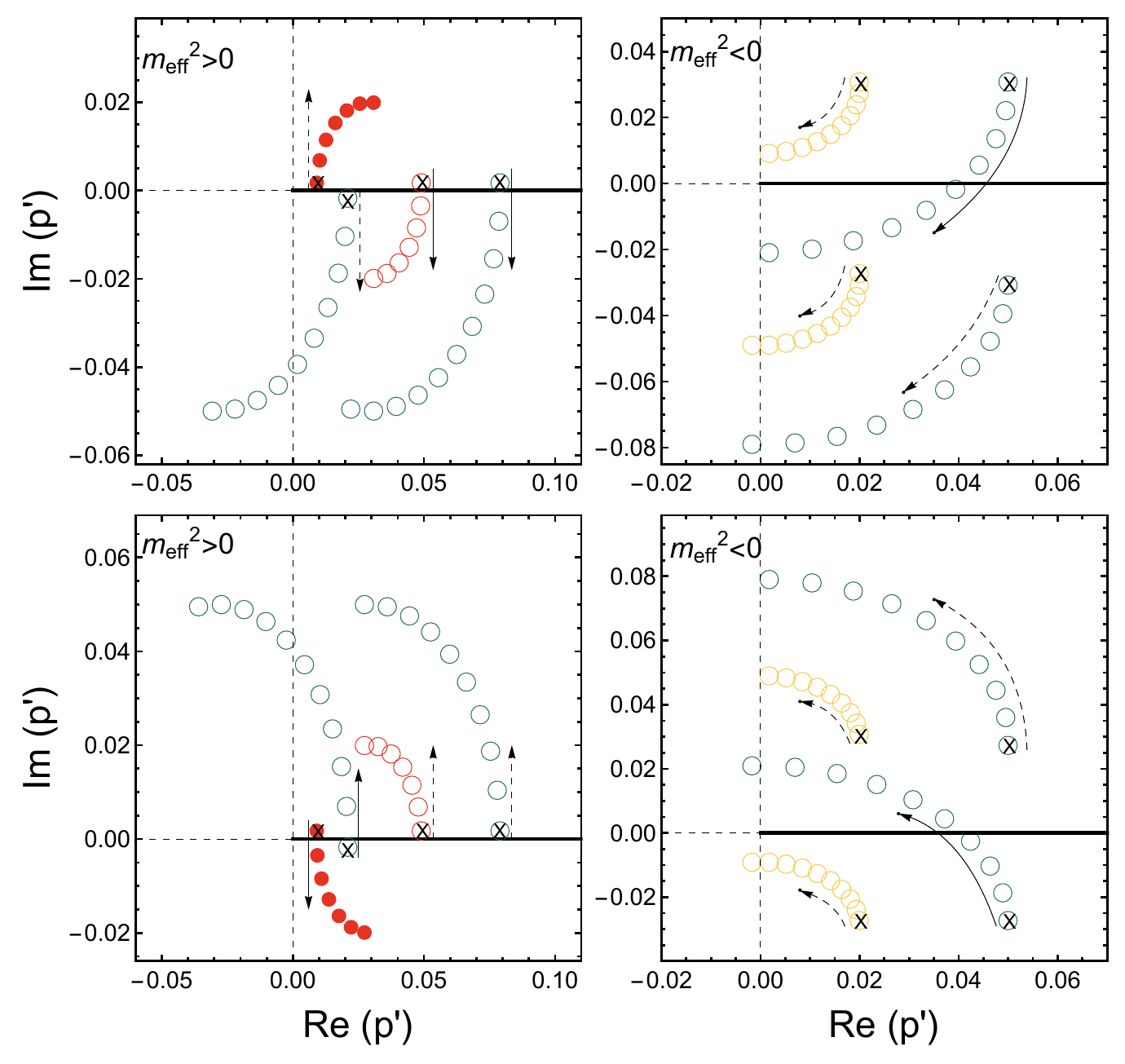}}
	\caption{The evolution of the branch points in the OPE potential with the three-body dynamics using complex scaling transformations: $p = \text{\large{$\kappa$}} \to \text{\large{$\kappa$}}e^{-i\phi}$ (upper plot) and $p = \text{\large{$\kappa$}} \to \text{\large{$\kappa$}}e^{i\phi}$ (lower plot). The red (yellow) and green points correspond to $\text{\large{$\kappa$}}=0.01$ and 0.1 GeV, respectively, and points with the cross mark denotes $\phi=0^{\circ}$. \label{fig4} }
\end{figure}

In Fig. \ref{fig4}, we show the evolution of the branch points of the OPE potential involving the three-body dynamics in the complex plane when performing a complex scaling transformation $p=\text{\large{$\kappa$}}\to \text{\large{$\kappa$}}e^{-i\phi}$ and $p=\text{\large{$\kappa$}}\to \text{\large{$\kappa$}}e^{i\phi}$. For the clockwise transformation $p=\text{\large{$\kappa$}}e^{-i\phi}$, the hollow circle branch points move to the lower half-plane and solid circle branch points move to the upper half-plane, and vice versa. From Fig. \ref{fig4}, it is evident that certain branch points will traverse the positive real axis, consequently altering the relative positioning of the branch points with respect to the integral path along the positive real axis. To maintain analyticity and continuity of the $T$ matrix when extending it to the complex plane, a complex scaling transformation should also be applied to the integral path in LSE. For the clockwise transformation, the complex scaled Lippmann-Schwinger equation (CSLSE) can be written as 
\begin{eqnarray}  &&T(\text{\large{$\kappa$}}e^{-i\phi},p^{\prime},k_0)\bm{=}V(\text{\large{$\kappa$}}e^{-i\phi},p^{\prime},k_0)+\int_0^{\infty}\frac{d^3\bm{q}}{(2\pi)^3}e^{-3i\phi} \nonumber \\
&&~~~~~~~V(\text{\large{$\kappa$}}e^{-i\phi},qe^{-i\phi},k_0)G_{\gamma}(qe^{-i\phi},k_0) T(qe^{-i\phi},p^{\prime},k_0), \label{eq11}
\end{eqnarray}
where $T(\text{\large{$\kappa$}}e^{-i\phi},p^{\prime},k_0)$ can be numerically solved through an iterative equation.
For the counterclosewise transformation $p=\text{\large{$\kappa$}}\to \text{\large{$\kappa$}}e^{i\phi}$, the corresponding analytical continuation will become complicated. 
For $\text{\large{$\kappa$}}<|m_{eff}|$ on the physical pion mass condition, the solid circle branch point will traverse the positive real axis when $p=\text{\large{$\kappa$}} \to \text{\large{$\kappa$}}e^{i\phi}$. The transformation of the integral path should be $q\to qe^{-i\phi}$ instead of $q\to qe^{i\phi}$ and the corresponding CSLSE is 
\begin{eqnarray}  &&T(\text{\large{$\kappa$}}e^{i\phi},p^{\prime},k_0)\bm{=}V(\text{\large{$\kappa$}}e^{i\phi},p^{\prime},k_0)+\int_0^{\infty}\frac{d^3\bm{q}}{(2\pi)^3}e^{-3i\phi} \nonumber \\
&&~~~~~~~V(\text{\large{$\kappa$}}e^{i\phi},qe^{-i\phi},k_0)G_{\gamma}(qe^{-i\phi},k_0) T(qe^{-i\phi},p^{\prime},k_0).
\end{eqnarray}
While $\text{\large{$\kappa$}}$ exceeds a critical point, the hollow circle branch point, originating from the negative real axis as shown in Fig. \ref{fig4}, can cross the positive real axis and move faster than the rotational transformation. As a result, a logical path transformation of $q\to qe^{-i(180-\phi)}$ should be carried out. The similar conclusion can also be reached for a large $\text{\large{$\kappa$}}$ on an unphysical pion mass condition.

Through the CSLSE approach, a half-on-shell $T$ matrix $T(pe^{i\theta},k_0,k_0)$ can be derived, which can be subsequently reintroduced into the integral in Eq. (\ref{eq10}) by setting $p=p^{\prime}=k_0$. This allows for the successful calculation of the physical on-shell $T$ matrix.
Of course, one of the most important deduction derived from the CSLSE is that the on-shell $T$ matrix below the energy threshold $T(i\text{\large{$\kappa$}},i\text{\large{$\kappa$}},i\text{\large{$\kappa$}})$ should be computed using the integral path near the negative imaginary axis, just as $i\text{\large{$\kappa$}}$ corresponds to $\phi=90^{\circ}$. This implies that a straightforward extension of $p \to i\text{\large{$\kappa$}}$ in the conventional LSE in Eq. (\ref{eq9}) may be inadequate for dealing with this special interaction involving the three-body dynamics. This insight will be substantiated through a practical example later.

\section{Applications and conclusion}

\begin{figure}[t]
\centerline{\includegraphics[width=8.3cm]{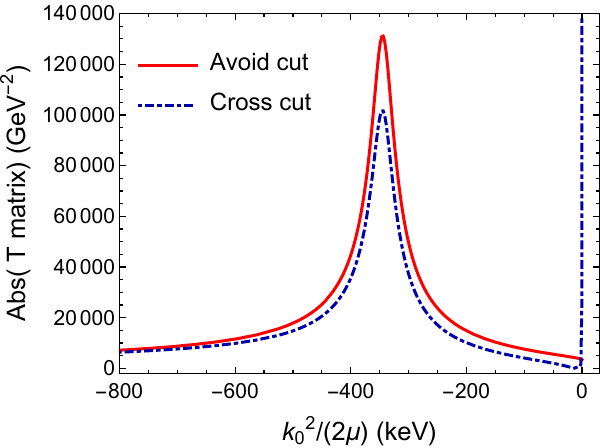}}
	\caption{The physical on-shell $T$ matrix of the isoscalar $DD^*$ scattering below the threshold. \label{fig5} }
\end{figure}

We now present a practical calculation of the isoscalar $DD^*$ scattering with the isospin symmetry by the CSLSE approach. In order to validate the reliability of our computed results, we focus on the pole information of $T_{cc}^{+}$, which was previously investigated in Ref. \cite{Lin:2022wmj} using the complex scaling method for Schrödinger equation. This method does not involve the branch cut complexities of the three-body dynamics and conveniently allows us to extract the dynamical pole position. The model parameters used to obtain the experimental $T_{cc}^+$ pole are listed below \cite{Lin:2022wmj}
\begin{eqnarray*}
&&C_t=-22.3~\mathrm{GeV}^{-2},~ \Lambda=0.5~\mathrm{GeV},~ m_{D}=1.867~\mathrm{GeV}, \\
&&m_{D^*}=2.009~\mathrm{GeV},~ m_{\pi}=0.139~\mathrm{GeV},~ g=0.65,~ \\&&f_{\pi}=0.086~\mathrm{GeV}, 
\end{eqnarray*}
where $\Lambda$ is a momentum cutoff. Since ChEFT only works at the small momentum regions and the CSLSE also requires the convergence at infinity of the integral, a Gaussian form factor is introduced to regularize the effective potential, which reads
\begin{eqnarray}
    \mathcal{F}(p,p^{\prime})=\mathrm{exp}[-\frac{p^2}{\Lambda^2}-\frac{p^{\prime2}}{\Lambda^2}].
\end{eqnarray}

The physical on-shell $T$ matrix of the isoscalar $DD^*$ scattering below the threshold are shown in Fig. \ref{fig5}, in which two different integral paths are adopted, i.e., the positive real axis and negative imaginary axis, which correspond to the path of crossing and not crossing the branch cut, respectively, as shown in Fig. \ref{fig2}.  In fact, there is flexibility to choose alternative rotation angles according to the Cauchy integral theorem, as long as the chosen paths consistently conform to either crossing or avoiding the branch cut. In the specific scenario, we have confirmed that the computation results are independent of the choice of the rotation angle. It is important to emphasize that in this context, where the branch cut arises from a logarithmic multivalued function, when the chosen integral path traverses the branch cut, an additional factor of $-2\pi i$ associated with the appropriate principal value must be taken into account.

From Fig. \ref{fig5}, it can be seen that the line shape distribution of the $T_{cc}^+$ state in the scattering amplitude can be accurately replicated, with the pole information matching that obtained through the complex scaling method of the Schrödinger equation \cite{Lin:2022wmj}. This provides compelling evidence for the validity of the CSLSE approach. Furthermore, we find that there is an unexpected divergency of the $T$ matrix at the two-body threshold for the integration scenario along the positive real axis, which is not caused by the precision of numerical calculation. A proof is presented as follows,  
\begin{eqnarray}
V_{\mathrm{OPE}}^{I=0}(i\text{\large{$\kappa$}},p)\propto \frac{\mathrm{Log}(1+(4\text{\large{$\kappa$}}pi)/(p^2-2\text{\large{$\kappa$}}pi-m_{eff}^2+i\epsilon))}{\text{\large{$\kappa$}}p},  
\end{eqnarray}
when $\text{\large{$\kappa$}}\sim\epsilon$ is a small quantity (close to the threshold). In the limit of $\text{\large{$\kappa$}} \to 0$, if the value of momentum $p$ traverses the branch cut, this expression can be further simplified as
\begin{eqnarray}
V_{\mathrm{OPE}}^{I=0}(i\text{\large{$\kappa$}},p)&\propto& \frac{-4i}{(m_{eff}^2-p^2-i\epsilon)}-\frac{2\pi i}{\text{\large{$\kappa$}}p}  \\
     &\sim& \frac{(2\pi i)}{\text{\large{$\kappa$}}p} \to \infty.
\end{eqnarray}
Obviously, this divergency at the $DD^*$ threshold should be unphysical. Hence, this point reinforces our conclusion that the accurate solution of the physical on-shell $T$ matrix below the threshold should be conducted within the CSLSE framework.

\begin{figure}[t]
\centerline{\includegraphics[width=8.3cm]{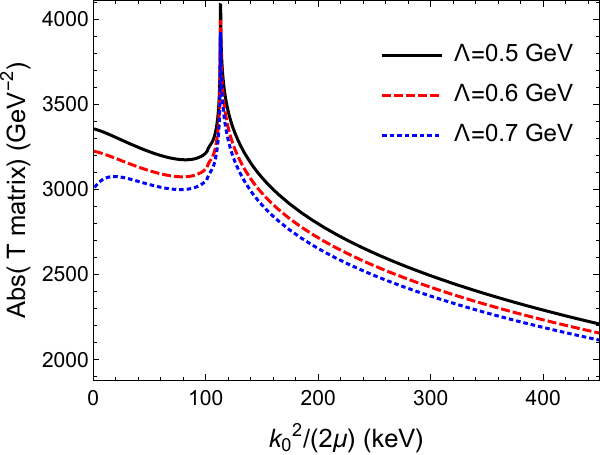}}
	\caption{The physical on-shell $T$ matrix of the isoscalar $DD^*$ scattering above the threshold. \label{fig6} }
\end{figure}

\begin{figure}[t]
\centerline{\includegraphics[width=8.0cm]{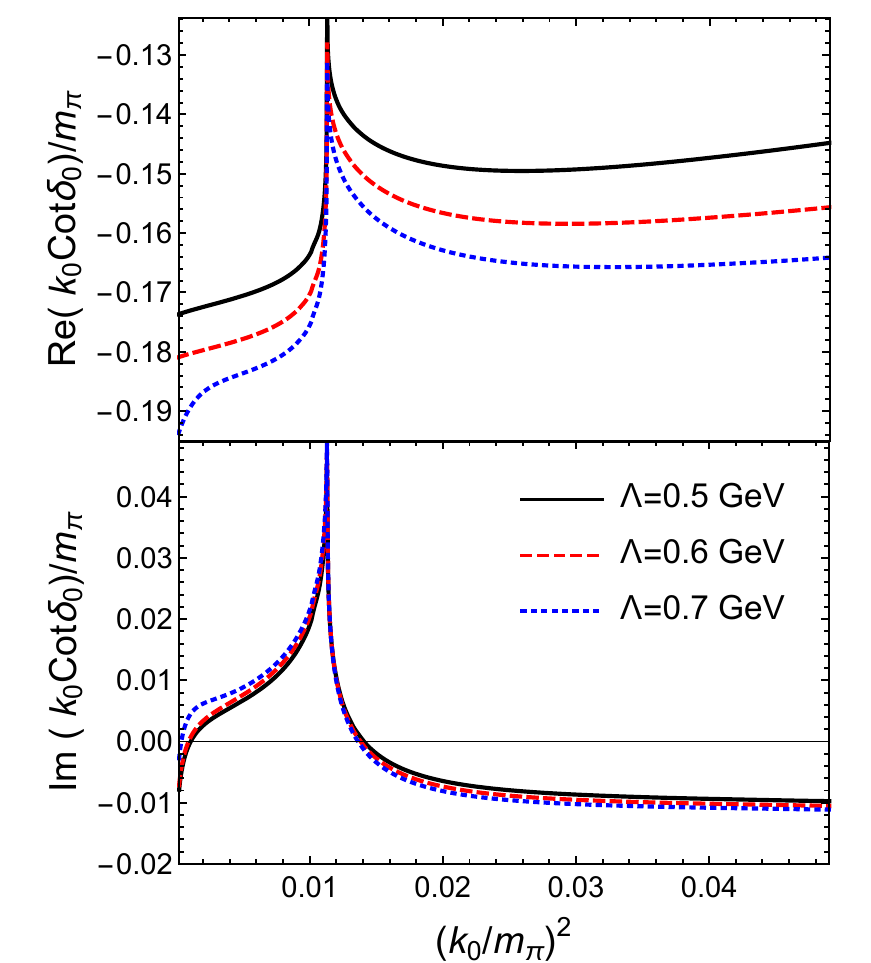}}
	\caption{The predicted scattering phase shifts $k_0\mathrm{cot}~\delta_0/m_{\pi}$ of the isoscalar $DD^*$ scattering.  \label{fig7} }
\end{figure}

Furthermore, we calculate the physical on-shell $T$ matrix of the isoscalar $DD^*$ scattering above the threshold, which is presented in Fig. \ref{fig6}. Very interestingly, the obtained $T$ matrix shows the existence of an extra new structure in addition to the reported $T_{cc}^+$ state. This new structure is due to the right-hand cut effect of the OPE potential involving the three-body dynamics. In order to further test our theoretical predictions, we also study the dependence on the cutoff parameter by changing $\Lambda$ between $0.5\sim0.7$ GeV. The existence of this structure is robust as shown in Fig. \ref{fig6}. The detection of this enhancement in experiments may necessitate the on-shell $DD^*$ beam scattering, a technology that could be explored in future high-energy experiments. In the current experimental setup involving the inclusive $DD^*$ production \cite{LHCb:2021vvq,LHCb:2021auc}, detecting this structure is challenging since the initial $DD^*$ is unnecessarily on-shell. From another perspective, the promising validation of this structure can be pursued through Lattice QCD simulations. We notice that the recent study of HAL QCD on $T_{cc}^+$ has extracted the $S$-wave scattering phase shifts of the $DD^*$ scattering at a nearly physical pion mass $m_{\pi}=146.4$ MeV \cite{Lyu:2023xro}. The $S$-wave scattering phase shift $\delta_0$ is directly related to the on-shell $T(k_0)$ matrix by
\begin{eqnarray}
   k_0~\mathrm{cot}~\delta_0=-\frac{8\pi^2}{\mu}T^{-1}(k_0)+ik_0. 
\end{eqnarray}
The phase shift $ k_0\mathrm{cot}~\delta_0/m_{\pi}$ obtained by Lattice QCD confirms a linear relationship with the variable $k_0^2/m_{\pi}^2$ \cite{Lyu:2023xro}.  This behavior can be ascribed to the lack of the three-body dynamics at an unphysical pion mass $m_{\pi}=146.4$ MeV used in their lattice simulations, where the trivial interaction enables the representation of the $T$ matrix through the effective range expansion, i.e., $T^{-1}\propto (1/a_0+1/2rk_0^2)$. In Fig. \ref{fig7}, we present the predicted scattering phase shifts when the three-body threshold dynamics is activated. Interestingly, the $ k_0\mathrm{cot}~\delta_0/m_{\pi}$ is no longer pure real numbers and the linear relation of its real part with the $k_0^2/m_{\pi}^2$ is seriously distorted.
This newly predicted structure arising from the right-hand cut is also manifested in both the real and imaginary parts of $ k_0\mathrm{cot}~\delta_0/m_{\pi}$, providing an opportunity for future verification through Lattice QCD simulations. 

In conclusion, we have developed a complex scaled Lippmann-Schwinger equation to address the complicated three-body threshold dynamics in low-energy hadron-hadron scatterings. The methodology is practically applied to the isoscalar $DD^*$ scattering process under a physical pion mass, and succeeds in reproducing the $T_{cc}^+$ structure, aligned with the pole derived from the complex scaling method in the Schrödinger equation framework. An essential finding from the CSLSE highlights the necessity of adjusting the integral contour of the loop momentum when investigating the physical on-shell $T$ matrix below the energy threshold. Moreover, by solving the on-shell $T$ matrix along the positive real axis of the momentum plane, a new structure stemming from a right-hand cut in the $DD^*$ mass spectrum is discovered for the first time. It is worth emphasizing that a similar novel structure is expected to exist in other systems such as $D\bar{D}^*, \Lambda_c\Sigma_c^{(*)}, \Lambda_c\bar{\Sigma}_c^{(*)}$, and more, serving as a distinctive indicator and presenting a valuable opportunity to explore the role of the three-body threshold dynamics in low-energy heavy-hadron-heavy-hadron interactions.

\section*{ACKNOWLEDGEMENTS}

This work is supported by the National Science Foundation of China under Grants No. 11975033, No. 12070131001 and No. 12147168. J.Z.W. is also supported by the National Postdoctoral Program for Innovative Talent. The authors thank Lu Meng, Yan-Ke Chen and Liang-Zhen Wen for helpful discussions.

\section{Appendix}

\subsection{A. An alternative scheme to deal with the singularities of the potential function in the integral}

In the domain of positive real momentum $p$, the partial-wave OPE potential involving the three-body dynamics $V(p,q)$ exhibits four singular points with the logarithmic divergence along the real axis of the momentum $q$ plane. In fact, such a potential singularity is actually integrable in the momentum interval of $q=0 \to q=\infty$. In a numerical algorithm, the integral can be discretized, i.e., 
\begin{eqnarray}
    \int V(p,q)f(q)dp=\sum_i V(p,q_i)f(q_i)\omega_i, 
\end{eqnarray}
where $f(q)$ stands for the integrand, and $\omega_i$ is the weight of the $i$-th quadrature point. Assuming that $V(p,q)$ is singular at points of positive real number $q=x_a$ and $x_b$, we can analytically integrate in the interval around the singularity, i.e., 
\begin{eqnarray}
  &&\int V(p,q)f(q)dq \to \sum_{q \neq x_a, x_b}V(p,q_i)f(q_i)\Delta q +f(x_a)   \nonumber \\
 &&\times\int_{x_a-\frac{\Delta q}{2}}^{x_a+\frac{\Delta q}{2}} V(p,q^{\prime})dp^{\prime}+f(x_b)\int_{x_b-\frac{\Delta q}{2}}^{x_b+\frac{\Delta q}{2}} V(p,q^{\prime})dp^{\prime},
\end{eqnarray}
where $\Delta q$ controls the numerical uncertainty. It is worth noting that the calculation efficiency of this algorithm applied in the conventional LSE is significantly lower than that of the CSLSE approach. In addition to the potential singularities, there is another singular point $q=k_0$ from the Green's function $G(q,k_0)=1/(q^2-k_0^2+i\epsilon)$, which can be dealt with a principle value integral and discontinuity.

\subsection{B. The cut structures of the three-body threshold dynamics with the kinetic energy terms of the heavy mesons }

\begin{figure}[t]
\centerline{\includegraphics[width=8.8cm]{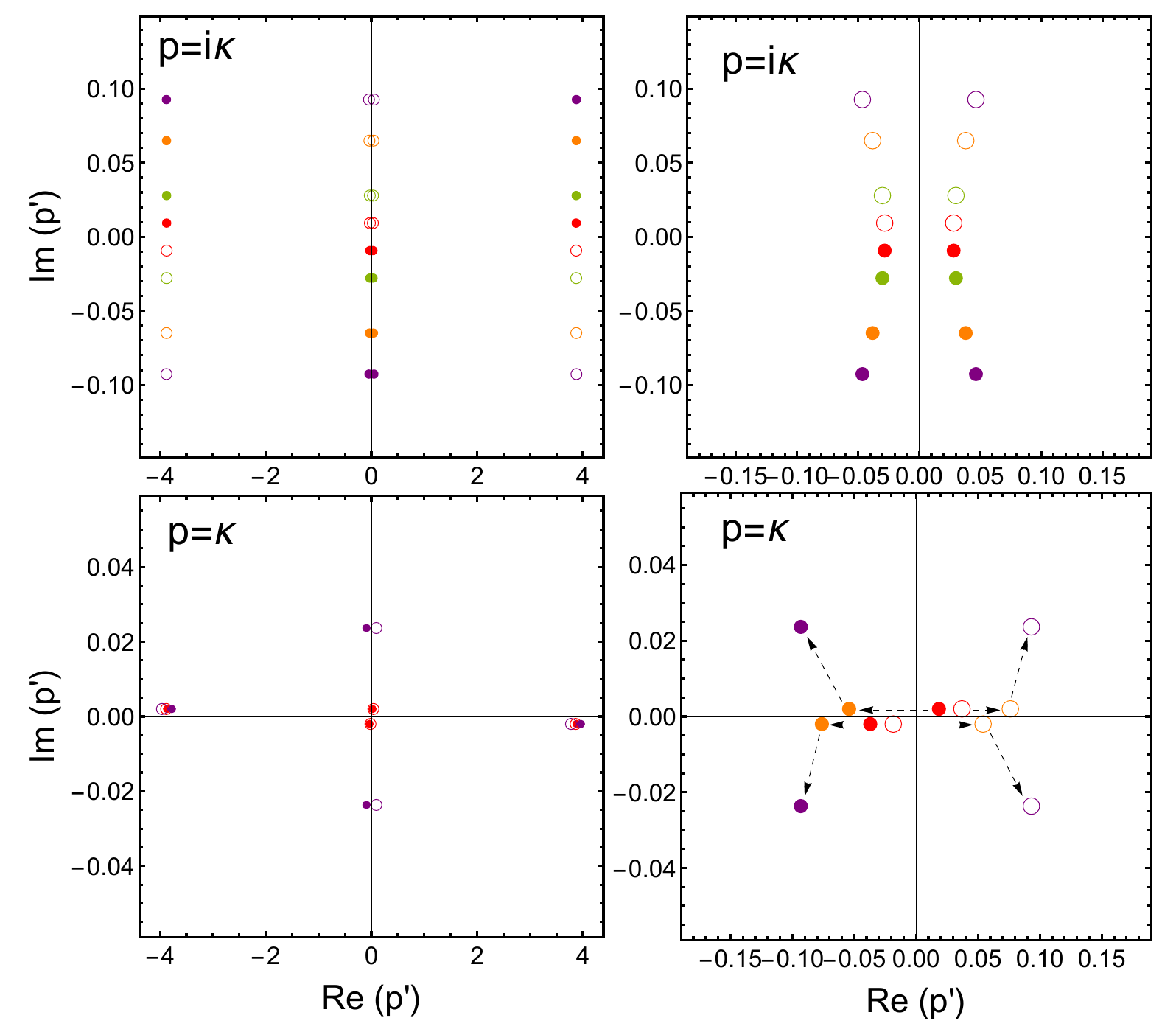}}
	\caption{The branch point distributions of the half-on-shell scattering amplitude of the three-body dynamics for $m_{eff}^2>0$ with the kinetic energy term of the heavy meson. The red, green, orange and purple singularities correspond to $\text{\large{$\kappa$}}=0.01$, 0.03, 0.07 and 0.1 GeV, respectively.  \label{fig8} }
\end{figure}

In the main text of this work, the discussions on the cut structures of the three-body threshold dynamics are based on the scenario of neglecting the kinetic energy terms of the heavy mesons in the pion propagator. When considering the kinetic energy terms of the heavy mesons, denoted as  $p^2/(2m_{D})$, the effective mass  $m_{eff}^2$ can be rewritten as follows:
\begin{eqnarray}
 m_{eff}^2=(E^{\prime}+m_{D^*}-m_{D}-(p^2+p^{\prime2})/(2m_D))^2-m_{\pi}^2,   \label{eq20}
\end{eqnarray}
which depends on the momenta of the initial and final states. In Fig. \ref{fig8}, we provide a direct comparison with Fig. \ref{fig2} by displaying the branch point distributions of the half-on-shell scattering amplitude for the three-body dynamics when considering the kinetic energy terms of the heavy mesons, where only the case with $m_{eff}^2>0$ is shown. The left half of the graph presents the full distribution, while the right half is an enlarged view of the branch points near the origin. Notably, under this scenario the number of branch points increases to eight. However, it is worth highlighting that these four extra branch points are situated near 4 GeV, significantly distant from the origin and, hence, considered unimportant due to the inapplicability of ChEFT at such large center-of-mass momenta.  Focusing on the four branch points around the origin, which correspond to the branch points presented in Fig. \ref{fig2}, we find that when $p=i\text{\large{$\kappa$}}$ is relatively small such as $\text{\large{$\kappa$}}=0.01$ GeV, the branch point positions remain almost unchanged compared to those derived from the three-body dynamics without the kinetic energy terms of the heavy mesons. As the momentum $p$ gradually increases from $p=i\text{\large{$\kappa$}}=i~0.01$ to $i~0.03$, $i~0.07$, and $i~0.1$ GeV, the branch points slightly shift to the upper right. Overall, the singularity positions remain similar to those in Fig. \ref{fig2}, where the kinetic energy terms of the heavy mesons are not considered.

An intriguing phenomenon occurs in the case with the positive real momentum $p=\text{\large{$\kappa$}}$. As mentioned earlier, their singularities in the $p^{\prime}$ plane lie along the real axis, presenting challenges for solving the dynamical equation due to their divergence on the integral path along the positive real axis. When $p$ is relatively small, its branch point behavior closely resembles that in Fig. \ref{fig2}. However, when momentum $p$ becomes sufficiently large, as indicated by Eq. (\ref{eq20}), there are  possibilities that $m_{eff}^2$ turns negative, causing the branch points to suddenly deviate from the real axis, and the corresponding circular branch cut undergoes significant changes. An upper limit for this transition point can be analytically estimated using
\begin{eqnarray}
    \text{\large{$\kappa$}}^{\mathrm{upper}}=\sqrt{2m_Dm_{D^*}-2m_D^2+2m_Dm_{\pi}},
\end{eqnarray}
which provides a rough estimate of the precise location of this transition point. This implies that in the large-momentum region of the scattering amplitude, the kinetic energy terms of the heavy mesons become crucial, and the resulting solution of the physical on-shell $T$ matrix above the energy threshold no longer encounters divergent singularities on the integral path along the positive real axis. A similar behavior is also found in the case involving an unphysical pion mass when solving for the physical on-shell $T$ matrix below the energy threshold.

\begin{figure}[t]
\centerline{\includegraphics[width=8.5cm]{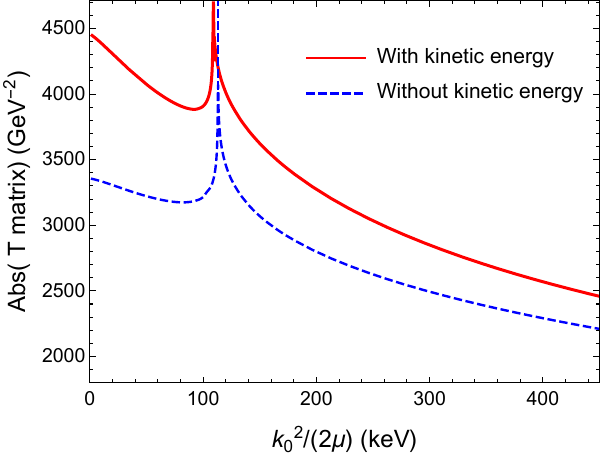}}
	\caption{ The comparison between the physical on-shell $T$ matrix of the isoscalar $DD^*$ scattering above the threshold with and without considering the kinetic energy terms of the heavy mesons. \label{fig9} }
\end{figure}

We have further investigated the physical on-shell $T$ matrix above the energy threshold within the context of the three-body dynamics, taking into account the kinetic energy terms of the heavy mesons and employing a momentum cutoff $\Lambda=0.5$ GeV. The corresponding results are presented in Fig. \ref{fig9}, further affirming the presence of the new structure originating from the right-hand cut.

\end{document}